# Possible traces of resonance signaling in the genome

Ivan Savelyev [1], Max Myakishev-Rempel [1,2,3]

1: Localized Therapeutics, San Diego, CA, USA

2: DNA Resonance Lab, San Diego, CA, USA

3: Transposon LLC, San Diego, CA, USA

**ABSTRACT**

**Although theories regarding the role of sequence-specific DNA resonance in biology have abounded for over 40 years, the published evidence for it is lacking. Here, the authors reasoned that for sustained resonance signaling, the number of oscillating DNA sequences per genome should be exceptionally high and that, therefore, genomic repeats of various sizes are good candidates for serving as resonators. Moreover, it was suggested that for the two DNA sequences to resonate, they do not necessarily have to be identical. Therefore, the existence of sequences differing in the primary sequence but having similar resonating sub-structures was proposed. It was hypothesized that such sequences, named HIDERs, would be enriched in the genomes of multicellular species. Specifically, it was hypothesized that delocalized electron clouds of purine-pyrimidine sequences could serve as the basis of HIDERs. The consequent computational genomic analysis confirmed the enrichment of purine-pyrimidine HIDERs in a few selected genomes of mammals, an insect, and a plant, compared to randomized sequence controls. Similarly, it was suggested that hypothetical delocalized proton clouds of the hydrogen bonds of multiple stacked bases could serve as sequence-dependent hydrogen-bond-based HIDERs. Similarly, the enrichment of such HIDERs was observed. It is suggested that these enrichments are the first evidence in support of sequence-specific resonance signaling in the genome.**

Ninety-seven years ago, Alexander Gurwitsch proposed the existence of a morphogenetic field that is created by the body and is responsible for developing and maintaining the shape of the body (A. Gurwitsch 1922). He and others demonstrated that biological organisms influence the development of each other at short distances and that some of this influence is blocked by optical filters, suggesting that the morphogenic field is of an electromagnetic nature (A. A. Gurwitsch 1988; Volodyaev and Beloussov 2015). In 1968, Frohlich predicted that in the presence of constant energy flux, cell and organelle membranes produce coherent waves in the millimeter-wave region, thus creating a coherent state and enabling electric wave signaling in living organisms (Frohlich 1988). In 1973, Miller and Web further proposed that it is DNA that is producing the morphogenic field and that the genomic code is directly sending and receiving the information from the morphogenic field (Miller and Webb 1973). The experiments verifying the existence of biological fields involve two samples such as cell culture aliquots in sealed quartz cuvettes separated by optical filters. When one of the aliquots is perturbed, the second one may catch a signal that is transferred non-chemically and is blocked by light-impermeable filters. Such effects are often referred to as "non-chemical cell-cell communication" and are reviewed in refs (Cifra, Fields, and Farhadi 2011; Scholkmann, Fels, and Cifra 2013; Trushin 2004; Xu et al. 2017). Burlakov experimentally demonstrated that the optical distortion by quartz retroreflectors of the field produced by fish embryos causes developmental abnormalities, thus confirming that the field is morphogenic and electromagnetic (Burkov et al. 2008; Burlakov et al. 2012).

Although the existence of the field and its morphogenic and electromagnetic nature have been demonstrated,

the involvement of DNA in its generation, proposed in 1973 by Muller and Webb, remains unproven. Many models for oscillations in DNA have been proposed that involve the movement of groups of atoms in DNA (referred to here as mechanical oscillations) (Scott 1985; Volkov and Kosevich 1987). The spectroscopic detection of coherent mechanical oscillations in DNA was reported to be in the THz range (Sajadi et al. 2011). Oscillations of delocalized pi-electron clouds of the base stack were experimentally observed in DNA (Xiang et al. 2015). We suggested that the base stack also supports natural oscillations of delocalized proton clouds of the hydrogen bonds (Savelyev et al. 2019). Furthermore, we suggested that electron and proton oscillations occur in a DNA sequence-dependent manner and provide the primary source for the formation of the morphogenic field. We also suggested that since electron and proton clouds have low mass and are located inside the base stack, they do not cause significant movement of the heavier DNA atoms and the surrounding water, thereby avoiding the thermal dissipation of energy. We suggested that, therefore, the electron and proton cloud oscillations in the base stack are a more likely source for the morphogenic field than the heavier atoms of DNA (Polesskaya et al. 2018; Savelyev et al. 2019).

We further proposed that electroacoustic resonances between similar DNA sequences form the basis of signaling within the genome and coordinate the function of the cell. We also suggested possible mechanisms by which these oscillations are channeled by the microtubules from one nucleus to another, forming an oscillation network of the body. This way, we transformed the idea of a diffuse morphogenic field into a model of the morphogenic field traveling between the nuclei via tunnels. This also explains how nature may avoid the dissipation of the electroacoustic signals in tissues (Savelyev et al. 2019).

We further implicated genomic repeats as primary candidate sequences to serve as resonators. We suggested that since the 300 base pair-long Alu repeat occurs 1.1 million times in each of our cells, it is the ideal candidate for serving as a resonator by the mere number of copies improving the sustainability of oscillations and reducing the dissipation of the signal. We also suggested that the primary positive function of genomic repeats such as telomeric, centromeric, simple repeats, and transposable elements is to support the resonance signaling in the genomes of complex organisms (Savelyev et al. 2019). Furthermore, we proposed (Savelyev et al. 2019) that this resonance signaling system is deliberately supported by the cells via the flux of ATP or other biochemical energy, in accordance with the Frohlich models (Fröhlich 1968). We suggested that similar DNA sequences resonate with each other, forming a resonating network within the nucleus, between the nuclei and across the organism. During this process, some of the repetitive sequences are energized by chemical processes, and their oscillations are transmitted along the base stack, causing oscillations in similar sequences. This way, conformational changes in the chromatin in one location lead to conformational changes in the chromatin of similar DNA sequences, allowing for resonance signaling within the nucleus, and across the organism, Fig. [Signal]. We suggested that this process is deliberate, developed by evolution for higher organisms and that the cell spends ATP and other types of chemical energy on supporting this resonance genomic signaling, Fig. [Signal]. This way, the chromatin mediates the information transfer between the electromagnetic and chemical signaling systems. During this signaling process, the resonance properties of the DNA sequences would provide specificity, while the ATP energy would allow for the amplification of electromagnetic resonance signals and their conversion to molecular signals. For example, oscillations in some Alu sequences may be induced by ATP-dependent chromatin remodeling factors, and these oscillations may be transmitted via the base stack to the second group of Alu elements. Via electromagnetic resonance, the Alu elements of the second group would begin resonance oscillation, which would be amplified by the ATP-dependent chromatin remodeling factors bound to them, causing chromatin opening and transcription. This proposed mechanism would explain why Alu elements are enriched in the gene promoters (Savelyev et al. 2019). The genome contains large numbers of greatly degraded copies of transposons, so it is likely that Alus in the promoters escaped the degradation because they served a useful function as transmitters and receivers of resonance signaling and thus participated in gene regulation.

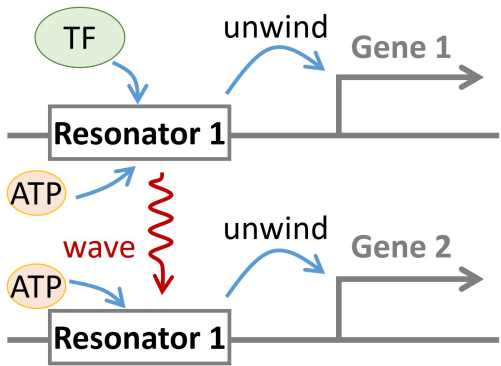

Fig. [Signal] Signal transduction via DNA resonance. A transcription factor binds to a DNA sequence of a Resonator 1 and initiates unwinding of the DNA, activation of nearby chromatin and transcription of Gene 1. Chemical ATP energy is used to generate electron and proton oscillations in Resonator 1. The oscillation signal is transferred by an electromagnetic wave to another copy of Resonator 1. ATP energy is used to receive the signal and activate nearby chromatin and transcription of Gene 2.

Although there are experimental demonstrations of morphogenic field effects, to our knowledge, the involvement of DNA in its formation is yet to be proven. The prediction of oscillation frequencies in DNA is not trivial since it is likely that DNA could support several oscillation modes, including those of mechanical oscillations and oscillations of electron and proton clouds. We suggest that sequence-specific oscillations in DNA can spread over a wide range of frequencies. Some insight might be obtained from electromagnetic frequencies used in physical therapy. Especially informative would be those frequencies, which produce effects at extremely low power, suggesting that they tap onto electromagnetic resonance signaling. Such frequencies are shown in Fig. [Spectrum].

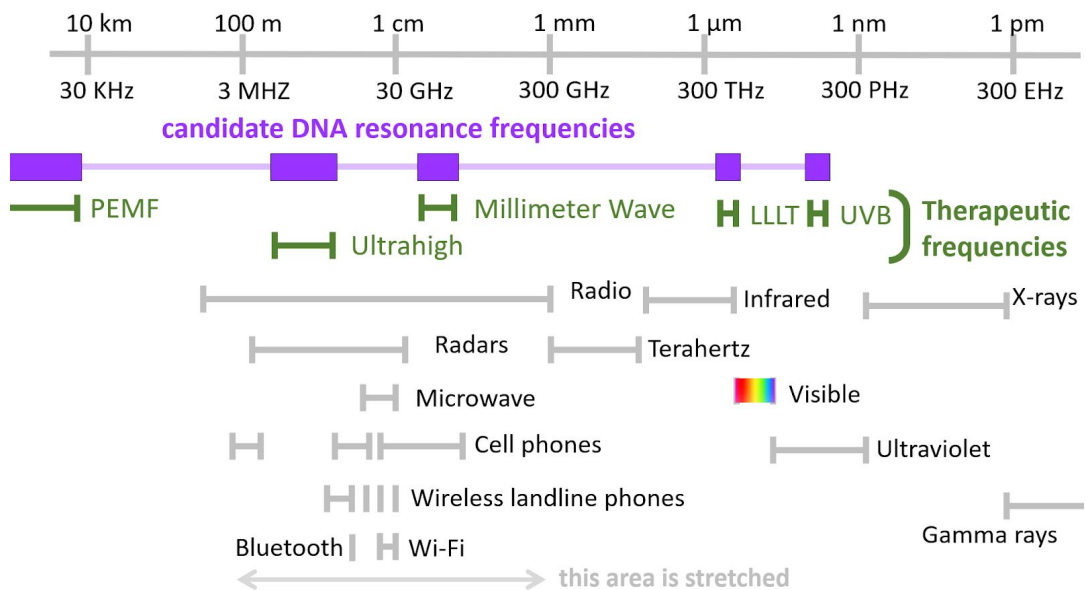

Fig. [Spectrum] Frequency ranges used for therapy. (LLLT – low-level light therapy, PEMF - pulsed electromagnetic field).

Specifically, the following therapeutic ranges of electromagnetic frequencies exhibit significant effects at low power, and thus are likely to be tapping existing signaling pathways: pulsed electromagnetic field therapy (Binder et al. 1984), ultra high-frequency therapy (Lushnikov et al. 2004), millimeter wave therapy (Usichenko, Ivashkivsky, and Gizhko 2003), Low-Level Light Therapy,LLLT (Bjordal et al. 2003), and UVB (Lowe et al. 1991). We suggest that these frequencies are good candidates for resonance oscillations in DNA. Since the frequency depends on the mass of the oscillator, shorter DNA repeats should oscillate at higher frequencies than the longer ones. Based primarily on these assumptions, we propose the following approximate prediction of resonance frequencies of the genomic repeats, **Table [Wavelengths]**. Note that the natural wavelength of the oscillator can be much larger than its size. Recently, radio communication saw the development of nanomechanical magnetoelectric (ME) antennas, which resonate at wavelengths 1000 times larger than their size (Nan et al. 2017; Shi et al. 2016). An additional conversion factor is that electromagnetic oscillations are coupled with the acoustic oscillations in biological tissues. Therefore, an electromagnetic wavelength from a therapeutics device may be converted to an acoustic wave in the tissue, thus shortening the wavelength approximately 200,000 times. Although the predictions in Table [Wavelengths] are preliminary and require experimental testing, they aim to illustrate the possible mechanistic connection between electromagnetic therapies and the proposed resonance genomic signaling.

**Table [Wavelengths]: A tentative prediction of resonance wavelengths of genomic repeats**

| Repeat unit length | | Periodic | Type | PEMF | UHF | MWT | LLLT | UVB |
|---|---|---|---|---|---|---|---|---|
| | | | light wavelength | 37km | 0.3m | 7mm | 800nm | 300nm |
| | | | sound wavelength | 186m | 1.5um | 30nm | 4nm | 1.5nm |
| 1 bp | 0.3 nm | y | simple | | | | | |
| 2 bp | 0.7 nm | y | simple | | | | | |
| 3 bp | 1.0 nm | y | simple | | | | | |
| 4 bp | 1.3 nm | y | simple | | | | | |
| 6 bp | 2.0 nm | y | telomeric | | | | | |
| 20 bp | 6.6 nm | n | Purine Hiders | | | | | |
| 20 bp | 6.6 nm | n | Strong Hiders | | | | | |
| 171 bp | 57 nm | y | centromeric | | | | | |
| 260 bp | 86 nm | n | MIR | | | | | |
| 300 bp | 100 nm | n | Alu | | | | | |
| 1000 bp | 332 nm | n | Mariner | | | | | |
| 6000 bp | 1992 nm | n | LINE1 | | | | | |

(UHF - ultra high frequency, MWT - millimeter wave therapy)

Since so far, to our knowledge, there has been no published evidence for resonance genomic signaling, we attempted to prove it computationally. Since we believe that the majority of repetitive sequences in the genome are involved in resonance signaling, we hypothesized that some of the unique (non-repetitive) sequences in the genome might have evolved to resonate with the genomic repeats. Accordingly, we hypothesized that it is not necessary for the unique sequence to be identical to the repeat, that for resonance, it might need to be only partially similar to the sequence of the repeat: for example, it is possible that some oscillations involve primarily the electron clouds of the aromatic rings (Savelyev et al. 2019). Therefore, only the purine-pyrimidine structure of the resonating sequences should be similar, while their primary sequences can be different. This simplification of the sequence from the primary sequence to the purine-pyrimidine sequence is hereafter referred to as the "Purine code." Similarly, for the oscillations primarily involving the hypothetical clouds of the delocalized protons of the hydrogen bonds in base pairs, only the patterns of these bonds should be similar, while the primary sequence can be different. This simplification of the sequence from primary to strong/weak (three bonds/two bonds per base pair), is hereafter referred to as the "Strong code." The recoding rules used here are listed in **Fig. [Codes]**.

| | | | |
|---|---|---|---|
| **Our recoding rules:** | | Purine | Pyrimidine |
| **Purine code** | A G → R - purines<br>C T → Y - pyrimidines | 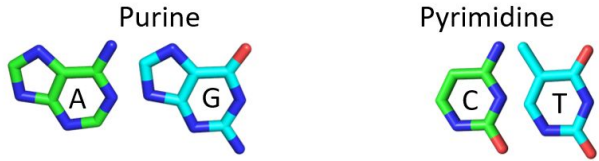 | |

| | | Strong | Weak |
|---|---|---|---|
| **Strong code** | G C → S - strong<br>A T → W - weak | 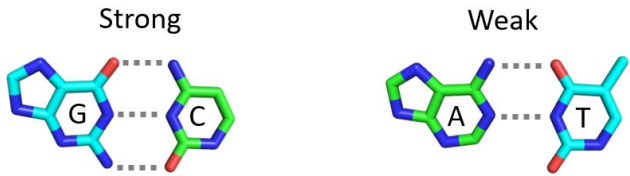 | |

| | | Amino | Keto |
|---|---|---|---|
| **Amino code** | A C → M - amino<br>G T → K - keto | 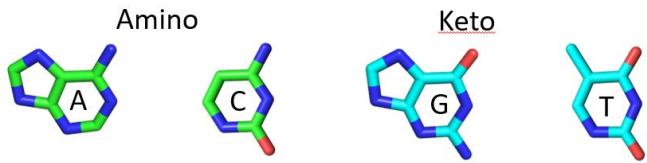 | |

| | | Thymine | Not thymine |
|---|---|---|---|
| **Thymine code** | T → T - thymine<br>A G C → V - not thymine | 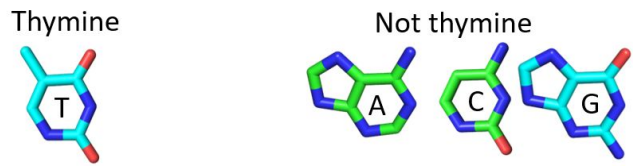 | |

*Fig. [Codes]. Recoding schemes used.*

Similarly, Amino and Thymine codes were used in the analysis. Therefore, we attempted to search for sequences that are unique (non-repetitive), but become similar to genomic repeats or each other after recoding (simplification). We will refer to them as HIDERs (Homologous If Decoded Elements, Repetitive). In accordance with the four recoding schemes, Table [Codes], four types of HIDERs were analyzed: Purine, Strong, Amine, and Thymine. On the primary sequence level, HIDERs are unique (non-repetitive) sequences, which are identical to each other after recoding, see for example Fig. [HIDERs].

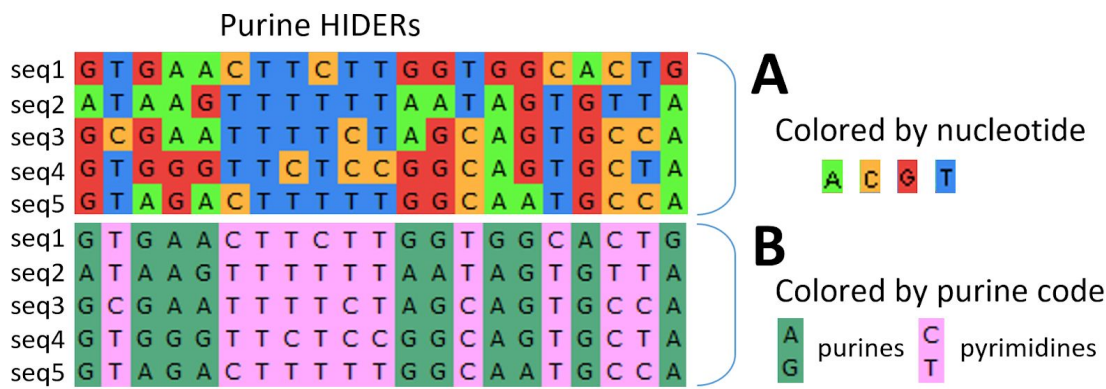

**Fig. [HIDERs]** HIDERs on sequence level. These HIDERs were picked from the sequence for their similarity in Purine code patterns. (A) and (B) contain exactly the same sequences seq1-seq5, the only difference is the coloring of the nucleotides. (A) nucleotides are colored by individual nucleotides. (B) Although the nucleotides are the same as in A, they are colored as purines vs pyrimidines, which we named "Purine code".

On the physical level, we expect HIDERs to be engaged in resonance signaling, and therefore, enriched in the genomes of complex organisms. One of the advantages of such a computational genomics approach is that it is agnostic to the exact physical mechanism of the resonance, allowing verification of its existence prior to the discovery of the mechanism. Once the existence of HIDERs is confirmed, their molecular structure may provide an insight into the modes of their resonance.

## Methods

The repeats were masked using RepeatMasker (http://repeatmasker.org/) followed by a heuristic removal of repeats using Ugene 1.32.0 (http://ugene.net/). Recoding was done as described in **Fig. [Codes]**. HIDERs were detected by searching for similar pairs of fragments in the recoded sequences using Ugene, analyzed in Google Sheets, and plotted with GraphPad Prism. Randomized sequences were used as controls, see Supplement. To retain the distribution of nucleotide densities along the sequence, randomization was done only on the unmasked parts of the sequence within each 20-nucleotide bin. The significance of HIDER enrichment was determined using the t-test. Vertebrate conservation scores were obtained from Vertebrate Multiz Alignment & Conservation (100 Species) Track from UCSC Genome Browser (Blanchette et al. 2004).

## Results

**HIDERs are enriched in genomes compared to randomized controls.**

We selected five species for analysis. In addition to humans, we chose mice, Drosophila, and Arabidopsis as typical model species and the dolphin as a highly developed aquatic mammal.

In each genome, four 90 kb pieces were selected at random. As expected in eukaryotes, the protein-coding sequences (exons) comprised a small fraction of the length. The repeats were masked. Randomized reference sequences were created from the original sequences. The Original and Randomized sequences were recoded, as presented in Fig. [Codes]. In each sequence, pairs of identical strings (HIDERs) longer than 19 bases were identified. The counts of HIDER pairs are shown in Fig. [Counts].

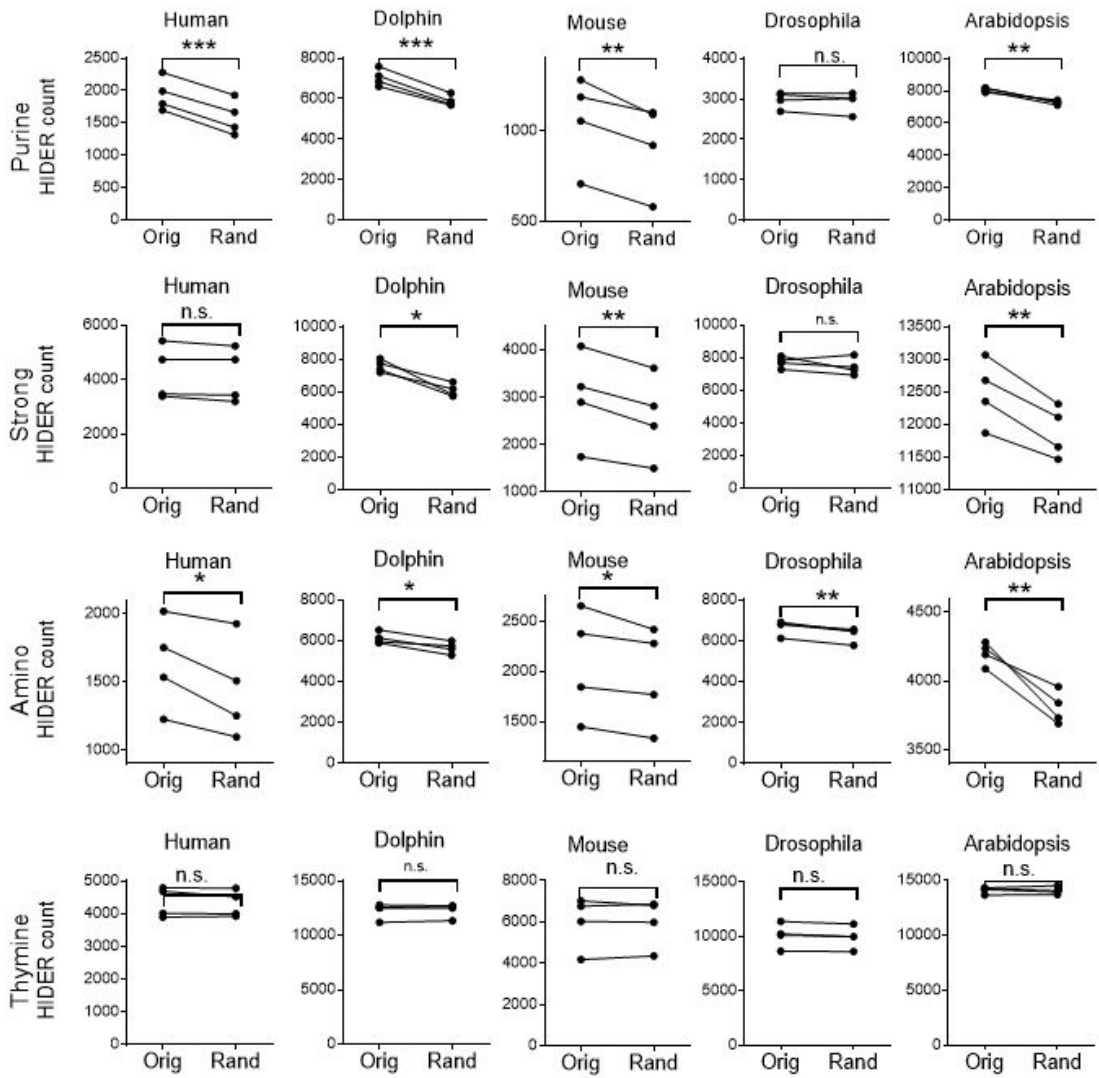

*Fig. [Enrichment] The enrichment of HIDER counts. Each line represents a 90kb genomic fragment. Note that the absolute counts of HIDERs naturally vary between species and chromosome positions, but the enrichment is characterized by the downward slope of the lines, from the original sequence (Orig) to its randomized version (Rand).*

The enrichment of HIDER counts in the original (Orig) over randomized (Rand) sequences from the same data is shown in Fig. [Summary].

$$\text{Enrichment} = \frac{\text{Count orig.} - \text{Count rand.}}{\text{Count orig.}} \%$$

| Code | Human | | Dolphin | | Mouse | | Drosophila | | Arabidopsis | |
|---|---|---|---|---|---|---|---|---|---|---|
| **Purine** | 19% | *** | 16% | *** | 14% | ** | 4% | n.s. | 10% | ** |
| **Strong** | 3% | n.s. | 20% | * | 14% | ** | 4% | n.s. | 5% | ** |
| **Amino** | 12% | * | 8% | * | 6% | * | 5% | ** | 9% | ** |
| **Thymine** | 1% | n.s. | 0% | n.s. | 0% | n.s. | 2% | n.s. | 0% | n.s. |

*Fig. [Summary]. Summary of enrichment of HIDER counts in the original sequences over randomized sequences. (\* - P < 0.05, \*\* - P<0.01, \*\*\* - P< 0.001, n.s. - nonsignificant)*

Among the five tested species, the highest enrichment of HIDERs was found in mammals and the lowest in Drosophila. Among the recoding schemes, the highest enrichment was found in the Purine code and the lowest in the Thymine code. The highest statistical significance was observed for the enrichment of Purine HIDERs in humans and dolphins.

In summary, statistically significant enrichment of HIDERs was observed in complex organisms.

**Tandem HIDERs.**

In addition to aperiodic hiders (such as shown in Fig. HIDERs), we inspected the enrichment of tandem (periodic) HIDERs, (such as shown in Fig. [Tandems]). Tandem HIDERs are repetitive in two dimensions: they are made of smaller repats (horizontal blue-orange periodicity in Fig. [Tandems], B)) as well as they exist in multiple copies in the genomic sequence (vertical similarity in Fig. [Tandems], B)). We suggest that the periodicity of Tandem HIDERs improves their coherence similarly to crystal oscillators used in quartz clocks. The computational analysis showed that Tandem HIDERs are significantly enriched in Strong code in 4 out of 5 complex organisms studied suggesting their positive function.

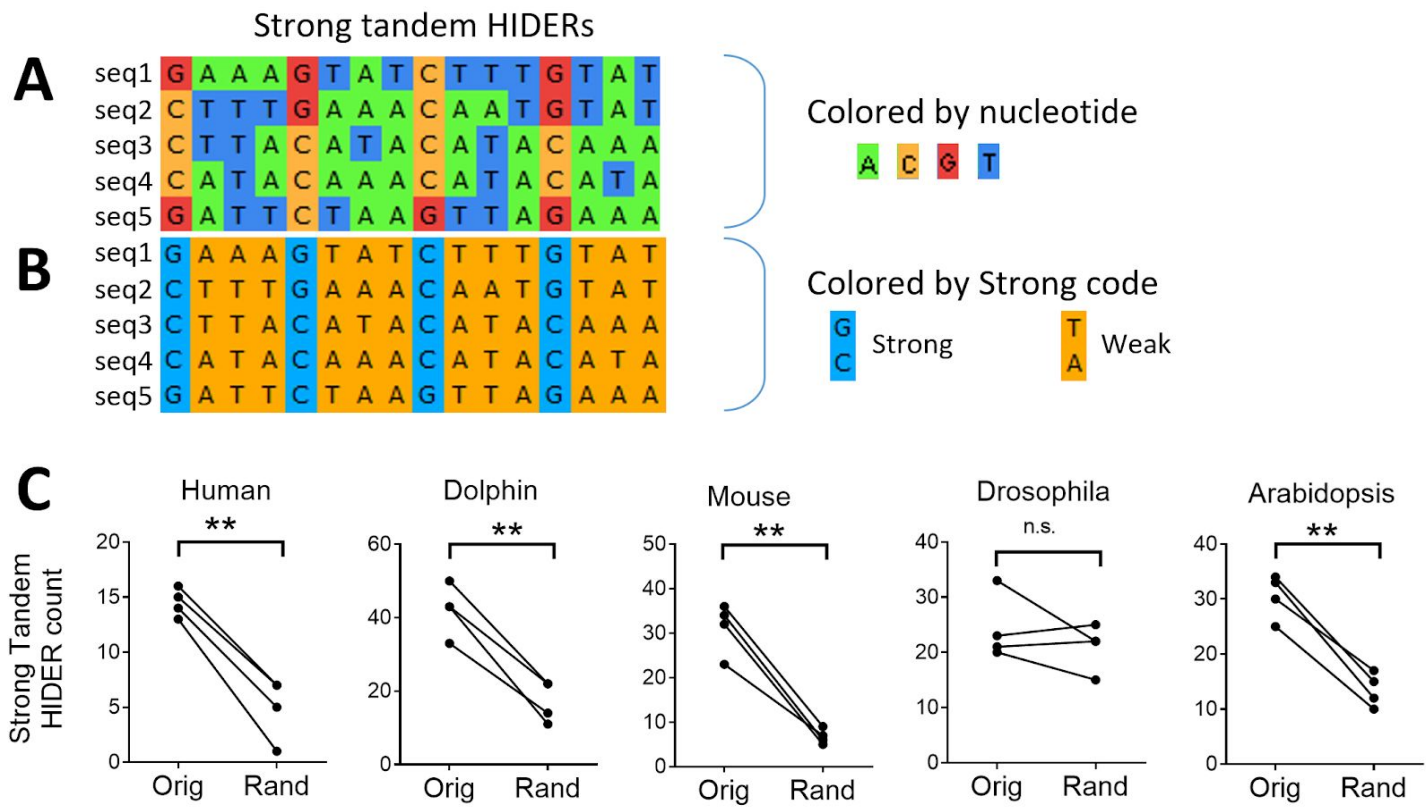

*Fig.[Tandems] Tandem Hiders. (A) Tandem HIDERs seq1-seq5 colored by individual nucleotides. (B) Same sequences seq1-seq5 colored by Purine code. (C) Enrichment of Strong Tandem HIDERs in original (Orig) compared to randomized sequence (Rand).*

We have also tested whether Strong Tandem HIDERs are located in conserved areas of the genome. Conservation across species is a measure that signifies functional importance of specific sequence regions. Typically, conserved sequence regions are responsible for morphogenesis and other functions common to many species. We observed with high statistical significance $P<0.0001$ that Strong Tandem HIDERs are located in conserved areas of the selected genomic region of Human genome, Fig. [Conservation]. The advantage of this observation is that instead of randomization, it uses average conservation per sequence are as a reference. Therefore evolutionary selection for Strong Tandem Hiders was confirmed by two largely independent computational methods: enrichment over a randomized sequence and localization to conserved areas.

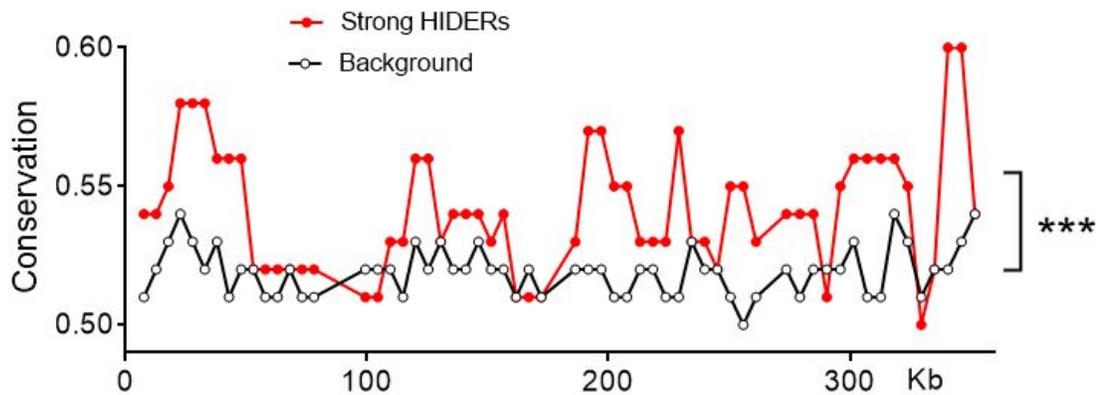

*Fig. [Conservation]. Conservation of Strong Tandem HIDERs. A 340 kilobase sequence was divided into 6kb bins. For each Strong Tandem HIDER, conservation across vertebrate species was calculated in a 300 base pair flanking region. Conservation of Strong Tandem HIDERs was averaged per bin (red). Conservation of all*

*sequence was averaged per bin (background, black). Strong Tandem HIDERs were 4% more conserved than the background, P<0.0001.*

## Discussion

As detailed in the Introduction, our initial motivation was to find sequence-dependent DNA resonators. We realized that for resonance to be sustained, the number of DNA resonators needs to be very high in each cell, in the order of millions of copies. Logically, so-called "junk DNA" made of repetitive elements of various sizes, is the primary candidate for harboring DNA resonators. We suggested that the key resonator in the human genome is the Alu element, which is represented by 1.1 million copies per cell. Then, we proposed that since DNA resonators ought to serve a function in coordinating the operation of the cell and the transfer of information between cells, resonator sequences should evolve to be enriched in the genome. Moreover, we hypothesized that even non-repetitive (unique) sequences might resonate with the repetitive sequences if they support similar modes of oscillation, that is, similar frequencies and patterns of electromagnetic oscillations. Then, we looked specifically for chemical structures in the DNA, which might support sequence-specific oscillations, and suggested that purine-pyrimidine patterns might be characterized by unique vibrational patterns. Specifically, we hypothesized that the pi-electrons of aromatic rings of sequential nucleobases might form a collective delocalized electron cloud. The shape and oscillation pattern of this cloud would be defined primarily by the purine-pyrimidine sequence. Therefore, DNA sequences, having a different primary sequence but common purine-pyrimidine patterns, might resonate. We called such sequences Purine HIDERs and suggested that they might be enriched in the genome. Here, we tested this hypothesis and confirmed the enrichment of the HIDERs in the selected mammal species and Arabidopsis but not in Drosophila, Fig. [Enrichment].

Similarly, we hypothesized that protons of hydrogen bonds of neighboring base pairs would form a delocalized proton cloud (a proton highway). This cloud would be prone to oscillations, and these oscillations would depend on the DNA sequence, specifically on the order in which base pairs with two hydrogen bonds (weak: A, T) and three hydrogen bonds (strong: C, G), respectively, occur in the DNA sequence. As above, we tested whether Strong HIDERs would be enriched in the genome. We observed a significant enrichment in the dolphins, the mice, and Arabidopsis, but not in humans or Drosophila, Fig. [Enrichment].

Note that both the Purine and Strong codes are simplifying the sequence from four symbols (A, C, G, T) to two symbols (purine/pyrimidine or strong/weak). Some information, including side radicals of the nucleobases, is lost, presumably allowing the HIDERs of different primary sequences to resonate with each other and likely with high-copy genomic repeats. Although we believe that high-copy genomic repeats are the primary resonators in the cell, we focused here on HIDERs since they allow us to test the DNA resonance hypothesis via computational genomics.

We are aware that in addition to the DNA resonance explanation, there are possible explanations for the observed enrichments that could involve traditional chemical causes. For example, it is possible that purines are more likely to mutate into each other than into pyrimidines, and vice versa. Therefore, repetitive sequences via the process of mutation might diverge in their primary sequence while retaining their purine-pyrimidine sequence, thus, effectively becoming HIDERs. Similarly, certain repeats might be the targets of transcription factors, which recognize their strong-weak pair sequence while ignoring actual bases. Consequently, certain repeats might diverge in evolution, producing Strong HIDERs. Currently available evolutionary base-substitution rates are not precise enough to enable the delineation of the chemical and resonance causes for the enrichment signal of HIDERs. Therefore, we hope that our results encourage further research and that the hypothesis of sequence-dependent DNA resonance signaling will be verified more conclusively.

One of the challenges in using computational genomics as a tool for testing hypotheses is the need for the Bonferroni correction in the case of multiple comparisons. To avoid multiple comparisons, we randomly selected the DNA fragments only once and did not optimize any analysis parameters. The selection and analysis of the data occurred only once. Moreover, to allow for testing by others, we selected only a few species and only a small part (approximately 3%) of each genome. This way, others could easily reproduce the observed enrichments on untouched data sets.

Since our research is focused on eukaryotes, it is important to consider chromatin structure and epigenetic modifications of DNA. Previously, we proposed a model of electric oscillations in tetranucleosomal units of chromatin (Polesskaya et al. 2018). A typical length of DNA wrapped around a tetranucleosome is 186x4=744 bp (base pairs) (Song et al. 2014). The structure of chromatin should strongly affect the oscillations of the tetranucleosomal resonators. Here, we focus on much smaller resonators, around 20 bp size (±10) which likely coexist with the larger, tetranucleosomal ones. In this study, the most significant enrichment was found using the Purine code. We have previously proposed a model of oscillations in a fragment in 20 bp size range (Polesskaya et al. 2018) which was based on purine-pyrimidine structure. Specifically, it was proposed that when purines are stacked on top of each other in the DNA double helix, their pi-electrons form a delocalized cloud; and the same for the pyrimidines stacked on top of each other. The currently obtained enrichment of purine hiders renders support to our previously proposed model (Polesskaya et al. 2018). The second highest enrichment result in the current study was produced using the Strong code. Similarly, this renders support to our previously proposed hypothesis that protons of the hydrogen bonds in the base stack form a delocalized proton cloud and the form of this cloud is dependent on the Strong code (Polesskaya et al. 2018). Since, importantly, the base stack is electrically insulated from the backbone, histones, and the nucleoplasm by hydrophobic repulsion (Sínanoĝlu and Abdulnur 1964; Arnold, Grodick, and Barton 2016), we believe that the chromatin structure will have a moderate effect on the oscillations in the DNA oscillators in 20 bp size range. The oscillation should be affected by bending and modifying the charge of the adjacent environment. Twenty bp corresponds to approximately 2 turns of the double helix, which is a quarter of nucleosome circumference. Binding of the DNA to a nucleosome bends the otherwise straight axis into a quarter-circle arc. This could affect the oscillatory properties of a 20 b.p. HIDER. For Purine code, chromatin binding and bending does not affect the basic purine-pyrimidine topology of the oscillator and thus it may retain the main oscillation properties while its chromatin environment is changed. The Strong code which is based on delocalized protons of hydrogen bonds will be more affected by the bending since bending might rearrange hydrogen bonds.

Also binding of the DNA to the nucleosome changes the charge environment of the base stack. According to our model, a cloud of delocalized charges (electrons and/or protons) is suspended and oscillates within the base stack. Electric charges surrounding the base stack should strongly affect these oscillations. The hydrophobic base stack is surrounded by a highly charged and hydrated DNA backbone and hydrated cations which neutralize the charge. We suggest that the electrically impermeable barrier between the base stack and the surrounding environment works as an electric condenser that couples the inner and the outer electric oscillations. Wrapping of DNA around the nucleosome will replace the ions and water with positively charged histones, thus strongly affecting the capacity of the condenser and thus affecting the frequency and quality of the oscillations.

The next important aspect of chromatin in eukaryotes is the methylation of cytosine. Unlike the DNA sequence, cytosine methylation varies from cell to cell and during the life of the cell. Based on chemical structure, cytosine methylation, should not affect the basic topology of purine-pyrimidine structure or hydrogen bonds, so the basic properties of Purine or Strong code-based resonators should not be strongly affected by the methylation. But cytosine methylation would certainly introduce subtle rearrangements of charges and hydrophobicity in the DNA chain and thus would affect the DNA resonators.

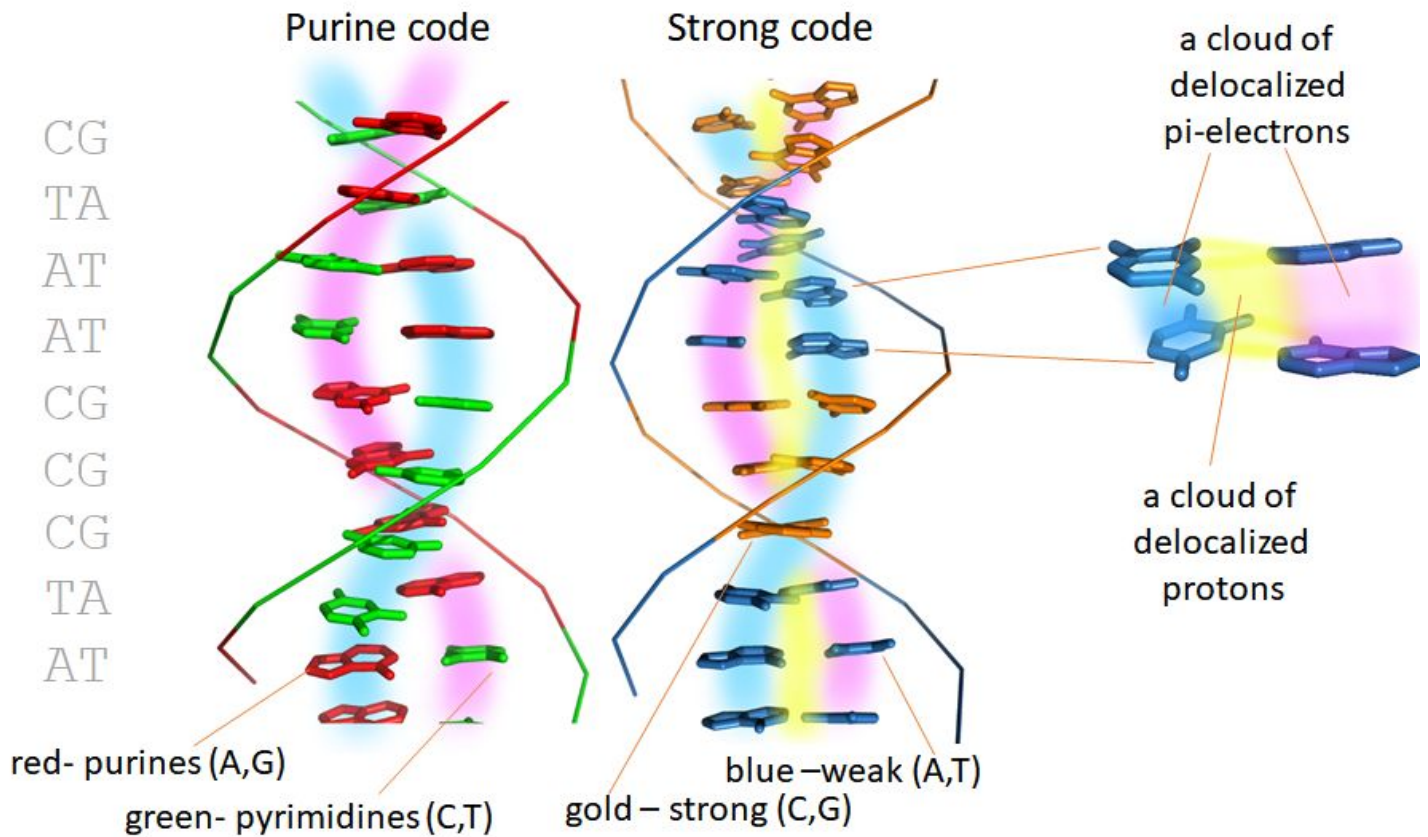

**Fig. [Clouds] Molecular structure of Purine and Strong codes, clouds of delocalized pi-electrons and delocalized hydrogen-bond protons.** The source sequence is on the left. The nucleotides are colored according to the Purine and Strong codes.

Our current molecular model of oscillators (Fig. [Clouds]) is not detailed enough to allow for calculation of oscillation frequencies for Purine and Strong hiders. Yet, based largely on the size, we tentatively estimated the oscillation wavelengths of the Purine and Strong hiders as shown in Table. [Wavelength]. Since the proton is 1836 times heavier than the electron, the wavelength of the Purine hiders (based on electron oscillations) should be shorter than the wavelength of the Strong hiders (based on proton oscillations), Table. [Wavelength]. Although the masses of electrons and protons are very different, possibly oscillations of their clouds are interacting and serve as signal frequency converters.

The overall logic of this study is shown in Fig. [Logic]. The morphogenic field hypothesis (i1) is supported by experiments of Gurwitch and Bulakov (i3). More experimentation and practical use will help its wider acceptance. To our knowledge, the hypothesis that the morphogenic field is generated by DNA (i2) has not been proven yet. Based on this hypothesis, we build a logical chain of sequential hypothetical assumptions i4-i10 and were able to confirm the latest hypothesis (i10) using the genome sequences of several species. The fact that HIDERs are enriched by evolution (i11) makes it very likely that HIDERs serve a positive function (i8). This provides the initial data-based evidence for the DNA resonance signaling (i9). The weakness of this proof (i9 by i11) is that there still is a possibility of non-resonance explanations of the observed enrichment of HIDERs discussed above. Yet, the attraction of this proof is that it is statistically significant, robust across locations and species and it is the first (although incomplete) evidence for the resonance signaling in the genome. Moreover, if DNA resonance signaling indeed exists, understanding it would have immediate value for biology and medicine.

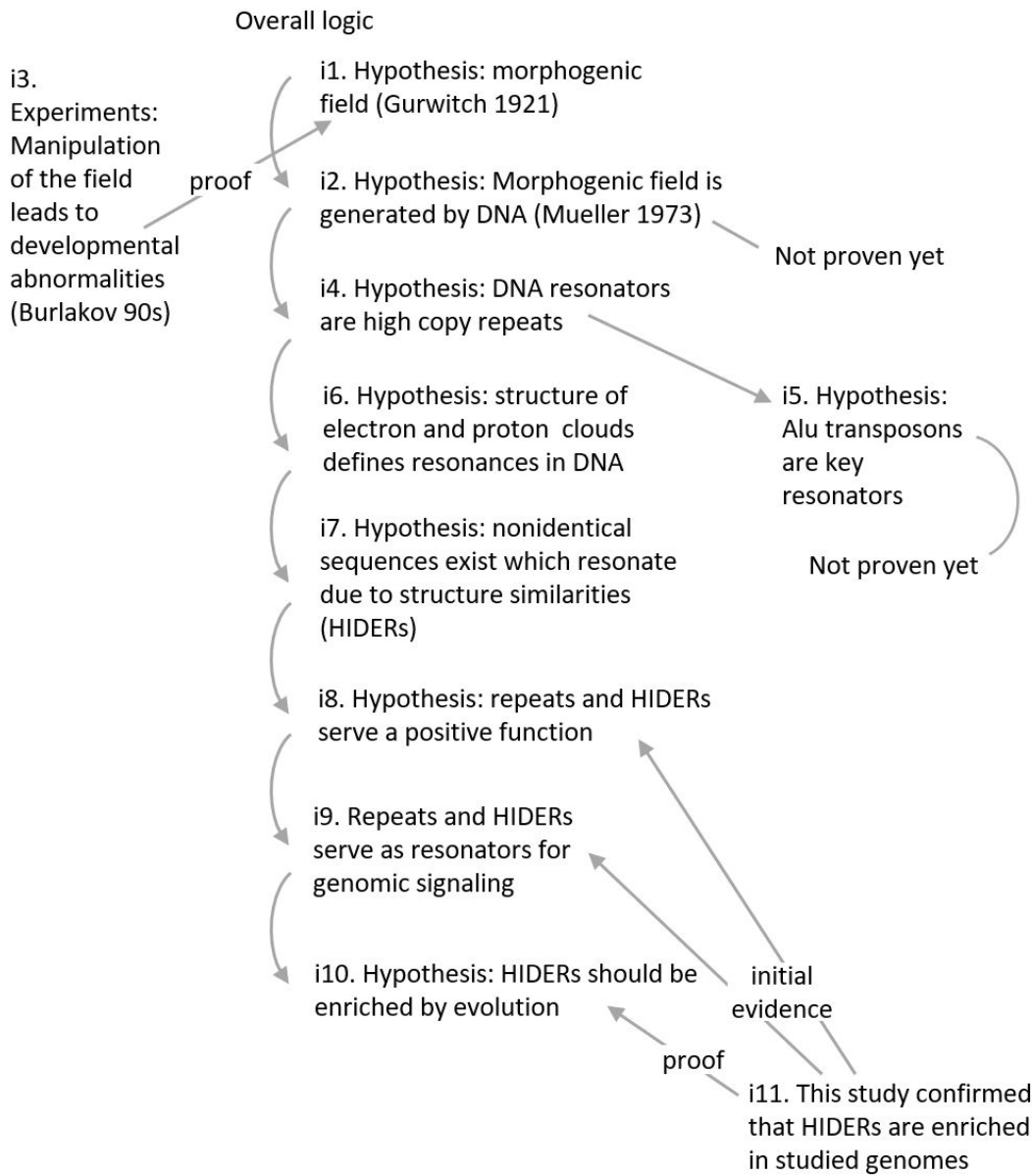

***Fig. [Logic] The overall logic of this study.***

Future research of resonance signaling in DNA may benefit both basic and applied science. Once proven and deciphered, resonance signaling in DNA could be utilized for a better understanding of genome regulation, diagnostics, and therapy. From molecular studies, it can be seen that often a single regulatory DNA sequence serves multiple purposes: in somatic cells, it is used for regulation of somatic functions and in the brain, it is used for brain functions. Similarly, we predict that DNA resonators such as HIDERs and genomic repeats are involved in somatic and neuronal signaling. Therefore, understanding the DNA resonance mechanism and signaling patterns may enable the development of medical applications for somatic diseases and brain disorders.

In summary, to our knowledge, this is the first, although indirect, evidence of DNA resonance in biology. However, although the obtained evidence is encouraging, more research is needed to verify the existence and the mechanistic details of DNA resonance. Computational modeling of the proposed electron and proton clouds of DNA sequences with the use of methods of quantum chemistry and structural biology could verify and substantiate the existence of such sequence-dependent resonating structures. Spectroscopic measurements could substantiate the proposed resonances between various sequences, including the ones highlighted by our analyses.

**Acknowledgments**.

We thank Oksana Polesskaya, Richard Alan Miller, Nelli Zyryanova and James E. Charles for discussion, and Martin Brockman and Juan Moran for programming. The work was funded by Max Myakishev-Rempel.

**References**

Arnold, Anna R., Michael A. Grodick, and Jacqueline K. Barton. 2016. "DNA Charge Transport: From Chemical Principles to the Cell." *Cell Chemical Biology* 23 (1): 183–97.

Binder, A., G. Parr, B. Hazleman, and S. Fitton-Jackson. 1984. "Pulsed Electromagnetic Field Therapy of Persistent Rotator Cuff Tendinitis. A Double-Blind Controlled Assessment." *The Lancet* 1 (8379): 695–98.

Bjordal, Jan M., Christian Couppé, Roberta T. Chow, Jan Tunér, and Elisabeth Anne Ljunggren. 2003. "A Systematic Review of Low Level Laser Therapy with Location-Specific Doses for Pain from Chronic Joint Disorders." *The Australian Journal of Physiotherapy* 49 (2): 107–16.

Blanchette, Mathieu, W. James Kent, Cathy Riemer, Laura Elnitski, Arian F. A. Smit, Krishna M. Roskin, Robert Baertsch, et al. 2004. "Aligning Multiple Genomic Sequences with the Threaded Blockset Aligner." *Genome Research* 14 (4): 708–15.

Burkov, V. D., A. B. Burlakov, S. V. Perminov, Y. S. Kapranov, and G. E. Kufal. 2008. "[Correction of Long Range Interaction Between Biological Objects Using Corner-Cube Reflectors]." *Biomedical Radioelectronics*, no. 8-9: 41–48.

Burlakov, A. B., Y. S. Kapranov, G. E. Kufal, and S. V. Perminov. 2012. "[About Possible Influence on Biological Object Electromagnetic Fields]." *Weak and Ultraweak Fields and Radiation in Biology and Medicine-Proceedings of IV International Congress*, 111–12.

Cifra, Michal, Jeremy Z. Fields, and Ashkan Farhadi. 2011. "Electromagnetic Cellular Interactions." *Progress in Biophysics and Molecular Biology* 105 (3): 223–46.

Frohlich, H. 1988. "Theoretical Physics and Biology." *Biological Coherence and Response to External Stimuli. Berlin: Springer-Verlag*, 1–24.

Fröhlich, Herbert. 1968. "Long-Range Coherence and Energy Storage in Biological Systems." *International Journal of Quantum Chemistry* 2 (5): 641–49.

Gurwitsch, A. A. 1988. "A Historical Review of the Problem of Mitogenetic Radiation." *Experientia* 44 (7): 545–50.

Gurwitsch, Alexander. 1922. "Über Den Begriff Des Embryonalen Feldes." *Wilhelm Roux' Archiv Fur Entwicklungsmechanik Der Organismen* 51 (1): 383–415.

Lowe, N. J., J. H. Prystowsky, T. Bourget, J. Edelstein, S. Nychay, and R. Armstrong. 1991. "Acitretin plus UVB Therapy for Psoriasis. Comparisons with Placebo plus UVB and Acitretin Alone." *Journal of the American Academy of Dermatology* 24 (4): 591–94.

Lushnikov, K. V., Yu V. Shumilina, V. S. Yakushina, A. B. Gapeev, V. B. Sadovnikov, and N. K. Chemeris. 2004. "Effects of Low-Intensity Ultrahigh Frequency Electromagnetic Radiation on Inflammatory Processes." *Bulletin of Experimental Biology and Medicine* 137 (4): 364–66.

Miller, Richard A., and Burt Webb. 1973. "Embryonic Holography: An Application of the Holographic Concept of Reality." *DNA Decipher Journal* 2 (2). http://www.dnadecipher.com/index.php/ddj/article/view/26.

Nan, Tianxiang, Hwaider Lin, Yuan Gao, Alexei Matyushov, Guoliang Yu, Huaihao Chen, Neville Sun, et al. 2017. "Acoustically Actuated Ultra-Compact NEMS Magnetoelectric Antennas." *Nature Communications* 8 (1): 296.

Polesskaya, Oksana, Vadim Guschin, Nikolai Kondratev, Irina Garanina, Olga Nazarenko, Nelli Zyryanova, Alexey Tovmash, et al. 2018. "On Possible Role of DNA Electrodynamics in Chromatin Regulation." *Progress in Biophysics and Molecular Biology* 30: 1e5.

Sajadi, Mohsen, Kristina E. Furse, Xin-Xing Zhang, Lars Dehmel, Sergey A. Kovalenko, Steven A. Corcelli, and Nikolaus P. Ernsting. 2011. "Detection of DNA--Ligand Binding Oscillations by Stokes-Shift Measurements." *Angewandte Chemie, International Edition* 50 (40): 9501–5.

Savelyev, Ivan V., Nelli V. Zyryanova, Oksana O. Polesskaya, and Max Myakishev-Rempel. 2019. "On The Existence of The DNA Resonance Code and Its Possible Mechanistic Connection to The Neural Code." *NeuroQuantology: An Interdisciplinary Journal of Neuroscience and Quantum Physics* 17 (2). https://doi.org/10.14704/nq.2019.17.2.1973.

Scholkmann, Felix, Daniel Fels, and Michal Cifra. 2013. "Non-Chemical and Non-Contact Cell-to-Cell Communication: A Short Review." *American Journal of Translational Research* 5 (6): 586–93.

Scott, A. C. 1985. "Soliton Oscillations in DNA." *Physical Review A: General Physics* 31 (5): 3518–19.

Shi, Y., M. Choi, Z. Li, G. Kim, Z. Foo, H. Kim, D. Wentzloff, and D. Blaauw. 2016. "26.7 A 10mm3 Syringe-Implantable near-Field Radio System on Glass Substrate." In *2016 IEEE International Solid-State Circuits Conference (ISSCC)*, 448–49.

Sínanoĝlu, O., and S. Abdulnur. 1964. "HYDROPHOBIC STACKING OF BASES AND THE SOLVENT DENATURATION OF DNA." *Photochemistry and Photobiology* 3 (4): 333–42.

Song, Feng, Ping Chen, Dapeng Sun, Mingzhu Wang, Liping Dong, Dan Liang, Rui-Ming Xu, Ping Zhu, and Guohong Li. 2014. "Cryo-EM Study of the Chromatin Fiber Reveals a Double Helix Twisted by Tetranucleosomal Units." *Science* 344 (6182): 376–80.

Trushin, Maxim V. 2004. "Distant Non-Chemical Communication in Various Biological Systems." *Rivista Di Biologia* 97 (3): 409–42.

Usichenko, Taras I., Olexiy I. Ivashkivsky, and Vasyl V. Gizhko. 2003. "Treatment of Rheumatoid Arthritis with Electromagnetic Millimeter Waves Applied to Acupuncture Points--a Randomized Double Blind Clinical Study." *Acupuncture & Electro-Therapeutics Research* 28 (1-2): 11–18.

Volkov, S. N., and A. M. Kosevich. 1987. "Conformation Oscillations of DNA." *Molekuliarnaia Biologiia* 21 (3): 797–806.

Volodyaev, Ilya, and Lev V. Beloussov. 2015. "Revisiting the Mitogenetic Effect of Ultra-Weak Photon Emission." *Frontiers in Physiology* 6 (September): 241.

Xiang, Limin, Julio L. Palma, Christopher Bruot, Vladimiro Mujica, Mark A. Ratner, and Nongjian Tao. 2015. "Intermediate Tunnelling–hopping Regime in DNA Charge Transport." *Nature Chemistry* 7 (February): 221.

Xu, Jingjing, Fan Yang, Danhong Han, and Shengyong Xu. 2017. "Wireless Communication in Biosystems." *arXiv [physics.bio-Ph]*. arXiv. http://arxiv.org/abs/1708.02467.

# SUPPLEMENT

**Sequences used for the analysis.**
Four original and four randomized sequences were investigated,

**Random sequence selection method**: To avoid the multiple comparison problem in the statistics, the selection of sequences was performed only once. Each sequence was 90Kb long. The selection was achieved using a simple algorithm, and the coordinates were predetermined using a simple rule. The assemblies and the coordinates were as follows:

Human
hg38_dna range = chr1: 100000000-100090000
hg38_dna range = chr1: 100090001-100180000
hg38_dna range = chr1: 100180001-100270000
hg38_dna range = chr1: 100270001-100360000

Dolphin
turTru2_dna range=JH472452:10000-100000

turTru2_dna range=JH472452:181000-271000  
turTru2_dna range=JH472452:309250-399250  
turTru2_dna range=JH472452:480250-570250  

Mouse  
mm10_dna range=chr3:32500000-32590000  
mm10_dna range=chr3:32590001-32680000  
mm10_dna range=chr3:32680001-32770000  
mm10_dna range=chr3:32770001-32860000  

Drosophila  
dm6_dna range=chr2L:200000-290000  
dm6_dna range=chr2L:400000-490000  
dm6_dna range=chr2L:800000-890000  
dm6_dna range=chr2L:1200000-1290000  

 Arabidopsis  
hub_329263_araTha1_dna range=chr3:400000-490000  
hub_329263_araTha1_dna range=chr3:600000-690000  
hub_329263_araTha1_dna range=chr3:1490000-1580000  
hub_329263_araTha1_dna range=chr3:1800000-1890000  

**Repeat masking**

Repeat masking was conducted in two steps. The original sequence was uploaded into the online RepeatMasker service (http://repeatmasker.org/cgi-bin/WEBRepeatMasker), and the repeats were masked with Ns. Then the sequence was masked by the Find Repeats algorithm of the UGENE program (Unipro UGENE http://ugene.net/).

**Search for HIDERs in the recoded sequence.**

The masked sequence was randomized as described in the Methods section. The original and randomized sequences were transformed into degenerate codes, as shown in Fig. [Codes]. Pairs of identical HIDERs longer than 19 pairs were identified using the Find Repeats algorithm of the UGENE program. The accuracy of the search was tested in part by verifying that the primary source sequences for the pairs of HIDERs were different as intended.

**Length dependence**

In Arabidopsis, Purine HIDERs demonstrated a positive correlation of the enrichment of HIDERs with their length, Fig. [Length]: the enrichment was higher for longer HIDERs. This suggests that longer HIDERs might be functional and, thus, preferentially selected during the process of evolution. Such correlation was less pronounced in the other species studied.

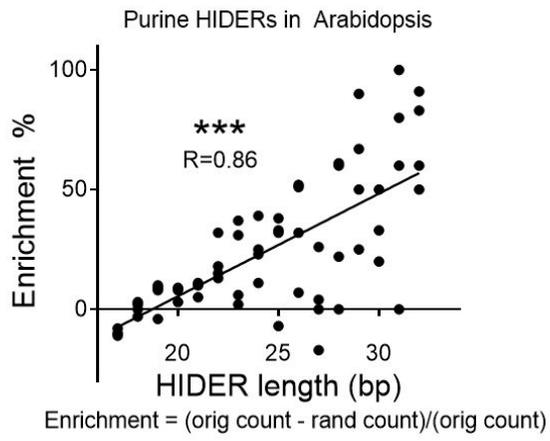

*Fig. [Length] Length dependence of Purine HIDER enrichment in Arabidopsis.*